\documentclass{article}

\usepackage{microtype}
\usepackage{graphicx}
\usepackage{subfigure}
\usepackage{booktabs} 
\usepackage{multicol}
\usepackage{multirow}
\usepackage[utf8]{inputenc}
\usepackage[T1]{fontenc}
\usepackage{graphicx} 
\usepackage{lipsum}
\usepackage{adjustbox}

\usepackage[accepted]{icml2021}

\icmltitlerunning{Who is responsible for adversarial defense?}

\begin{document}

\twocolumn[
\icmltitle{Who is Responsible for Adversarial Defense?}



\icmlsetsymbol{equal}{*}

\begin{icmlauthorlist}
\icmlauthor{Kishor Datta Gupta}{equal,to}
\icmlauthor{Dipankar Dasgupta}{equal,to}
\end{icmlauthorlist}

\icmlaffiliation{to}{University of Memphis, Memphis, TN, USA}

\icmlcorrespondingauthor{Kishor Datta Gupta}{kdgupta87@gmail.com}

\icmlkeywords{Machine Learning, ICML}

\vskip 0.3in
]



\printAffiliationsAndNotice{\icmlEqualContribution} 

\begin{abstract}
We have seen a surge in research aims toward adversarial attacks and defenses in AI/ML systems. While it is crucial to formulate new attack methods and devise novel defense strategies for robustness, it is also imperative to recognize who is responsible for implementing, validating and justifying the necessity of these defenses. In particular, which components of the system are vulnerable to what type of adversarial attacks, and the expertise needed to realize the severity of adversarial attacks. Also how to evaluate and address the adversarial  challenges in order to recommend defense strategies for different applications. This paper opened a discussion on who should examine and implement the adversarial defenses and the reason behind such efforts.
\end{abstract}

\section{Introduction}
\label{submission}

Several studies have been conducted on the challenges of ML model deployment \cite{ferguson2019standardized}. Primarily these study discusses challenges on data scaling, testing, evaluating model, etc\cite{baier2019challenges}. AIOPS (AI-operators) terms were introduced by \cite{dang2019aiops} where the role of developers, program managers, support engineers, site reliability engineers are mentioned for handling AI model deployment challenges. Still, it did not elaborated adversarial attacks. Researcher Paleyes et al. \cite{paleyes2020challenges} provide a detailed case study of an ML model deployment challenge and also extend discussion on the adversarial challenge in ML deployment. However, this study is also limited in terms of adversarial attacks and ignored where the adversarial attack can formulate. Paleyes mainly consider Devops (or MlOps?) for deployment issue, which are questionable from the security and robustness perspectives. This paper argues that other engineering roles need to be considered from an adversarial attack perspective. Based on our literature review, no significant research has been done on the ML model security in design and production, indicating important need in this domain.

There exist well-established positions for data scientists/engineers with proper job description who are responsible for  training and testing of ML models. Moreover, after deploying an ML model in a system either distributed/centralized or cloud/IoT environment will also require monitoring and inspection. As system infrastructure engineers share the deployment responsibility, the security of AI/ML model can be viewed as part of the system security and may fall in the cybersecurity expertise area of expertise. However, adversarial defenses are diverse and standards  for AI/ML risk management and associated tools are not available yet. It can be argued that defense against adversarial attacks should have been shared between all concerned parties and need to be mitigated with proper knowledge to be effective.
In this paper, we tried to introduce an discussion about the responsibility of adversarial defenses of AI-based systems.

\section{Preliminaries}

\subsection{Adversarial Attacks and their Strange Nature}
\label{th}
Adversarial attacks (AA) manipulate input data by adding perturbs/noises in various malicious ways. It is to be noted that the explanation behind AA's existence has no definitive answer. Some asserts the reason is the non-linearity of ML\cite{szegedy2013intriguing}, but some \cite{goodfellow2014explaining} proclaims it is too much linearity in ML models. Another argument by \cite{tanay2016boundary}, offered a tilted boundary theory and emphasized that it is never feasible to fit a model entirely to eliminate AAs. MIT researchers stated that all adversarial features are not noise as these data cannot be properly classified because human sensors are not advanced adequate to incorporate a class for those data. Though, this reasoning is contradicted by other researchers\cite{ilyas2019adversarial}.

As per NIST definition, we can define AAs into three basic type as\cite{dasgupta2020machine} :
\begin{enumerate}
\small
    \item Poisoning attack: By changing ML training data. \textbf{Happens in training time}.
    \item Evasion attack:  Inputs change in a way that they miss-classify to another random or targeted class\cite{dasgupta2020machine}.\textbf{Happens in production. }
    \item Trojan AI attack: In model's architecture changes in a way it miss-classifies the input. It can \textbf{happen in ML model setup/transfer to system/runtime \cite{gu2017badnets}.}
\end{enumerate}

As a poisoning attack is not an attack on deployed ML systems, we will not consider it in this paper. This paper will focus on the attacks carried out in the deployed system.

Ian Goodfellow introduced the first gradient sign method (FGSM). Another perturb method known as the Basic iterative method (BIM) works as an extension of FGSM. Madry added a random start before BIM named that PGD. DeepFool attack types generated by iteratively linearizing the decision boundary. Carlini-Wagner (CW)introduced adversarial examples with the slightest perturbations. The local search attack was introduced in 2016. In 2018, the EAD attack, which is elastic-net regularized optimization-based attack , decoupling based attack, Sparse attack based on multiple perturbations introduced\cite{dasgupta2020machine}.

The adversarial patch attacks \cite{brown2017adversarial} are distinct from the other attack types, as humans can easily recognize noises. Some variations of these attacks are LAVAN\cite{karmon2018lavan} and DPATCH\cite{liu2018dpatch}.

Another important nature of AAs are transferability \cite{zhou2018transferable} i.e. adversarial input generated for a ML model can work for different ML model.

In 2017, \cite{kurakin2016adversarial} pointed that in the digital version and printed version effectiveness of the adversarial method's drop. They attempted to defend their argument with FGSM, BIM, and other iterative methods. Also, \cite{lu2017no,nguyen2019detecting,gupta2020applicability} experimented with FGSM, BIM, and LBFGS and confirmed the drop of AA effectiveness rate up to 100\%, thus nullifying these attacks. It has proven that empirically and theoretically that AAs with more subtle noise are more vulnerable for physical existence. 

Researchers \cite{prakash2018deflecting,akhtar2018adversarial,gupta2020determining} show adversarial and clean images have a significant differences in their noise value which are easily identifiable for attacks such as FGSM, BIM etc.

\subsection{Limitations of Adversarial Defenses}
\label{ch}
Major defense techniques are Adversarial Training ,Distillation (train model twice), Pre-Processing Using PCA, low-pass filtering, data compression, soft threshold ,Model Structure modification, Network verification method  ,Ensemble multiple defense method \cite{dasgupta2020machine} and several benchmarks ~\cite{carlini2019evaluating, athalye2018robustness, carlini2017magnet, carlini2017adversarial,carlini2019lessons,tramer2020adaptive}  are available to evaluate defense systems, but these evaluations concentrated on how many distinct types of attacks a defense system can defend, particularly prioritized effectiveness against adaptive attacks.  

Adversarial training diminishes the ML model's accuracy and can make the ML model more exposed to generalization \cite{raghunathan2019adversarial}. Another disadvantage of Adversarial training based defense techniques is that we need to retrain the model whenever some new attack samples are discovered. It will be hard to update all deployed ML models.
Most pre-processing techniques reduce the adversarial effect before sending it to the ML model. The major drawback of these techniques is that their processing techniques are static; they do not evolve alongside the attack. 
Distillation techniques work by combining the double model, and the second model uses the first model knowledge to improve accuracy. The black-box attack's recent improvement makes this out-of-date defense \cite{chakraborty2018adversarial}. The strong transfer-potential of adversarial samples across neural network models \cite{papernot2016distillation} is the main reason for this method's collapse. It is not robust as simplistic variation in a neural network can make the system exposed to attacks \cite{carlini2016defensive}. 
\cite{he2017adversarial} concluded that combining/ensemble weak defenses does not automatically improve a system's robustness. Also, the ensemble technique remains static and vulnerable to a new attack. 
Feature squeezing \cite{xu2017feature} method reduces the data, and it reduces the accuracy of the ML model. There is no such reduction in actual model accuracy in our proposed solution. \cite{samangouei2018defense} proposed a mechanism to leverage the power of Generative Adversarial Networks to decrease adversarial perturbations' efficiency. The GAN efficiency depends on the GAN training, which is computationally complicated.

\section{Should we care about Adversarial Attacks?}

 \begin{figure*}[ht]
\centering
    \includegraphics[width=1\linewidth]{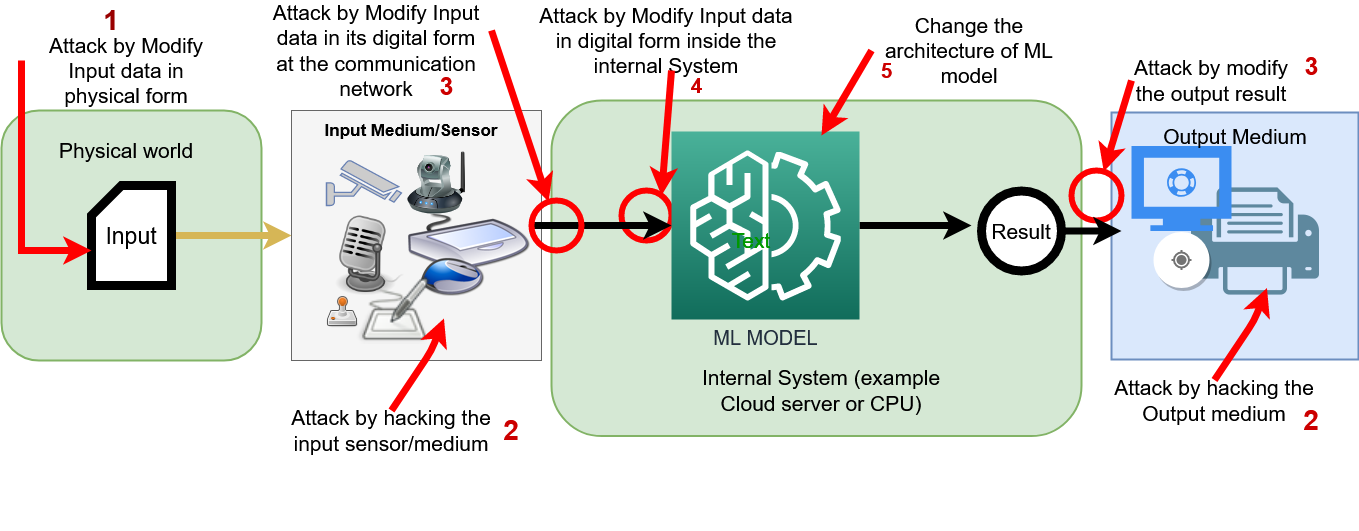}
        \caption{Generalized adversarial attack points on ML model}
        \label{fig:radar}
\end{figure*}

From the discussion in section \ref{th}, we can see evasion-based AAs are difficult to formulate in the real world. Environmental factors can easily cancel out adversarial noises. The noises which are not canceled out are also very easily detected by common filters. Most of the defense technique has 99\% accuracy rate of adversarial detection for high noise AAs ( example: FGSM, JSMA, BIM). So, this question can simply arise that do we need to care about adversarial attacks?

To answer this question, let's observe where an AA can be formulated. In the figure \ref{fig:radar}, we illustrated general points where adversarial manipulation could happen on a system where an ML service is deployed. We will consider the ML model as a black box. We marked the attack points by number. 

First, when the input is on the physical world, an adversary can change the input as a famous adversarial example of changing stop sign to speed limit sign conversion. Here, we have to note that this attack needs a higher physical degree of modification. Hence, if the noise amount is low, there is a high chance that environmental factors will nullify the attacks based on our literature review. 

The second point is primarily I/O layers. For example, if someone changes one of the sensors so that sensors can add adversarial noises while converting physical input data into a digital form. One prominent example is putting an adversarial patch sticker on the CC cam lens to manipulate ML output\cite{li2019adversarial}. This attack can also happen when an output action was presented or happening.

The third point is where data turned to a digital format (binary/floating/etc.) and sending to the ML system or output receiving from ML (Assuming the ML system is on a cloud server or some computational hardware). We can say it's more of a communication channel. It could argue that if adversarial can control this channel, intruder can modify the data going to the ML or coming from ML. Low perturb-based attacks can occur here as no environmental factors will diminish the noise. However, we can also say that if someone can control this network, intruder might not need to add adversarial noises; intruder can directly add his desired class input for his expecting output. Also, if intruder can modify the output result no need for adversarial perturb noises. Yet, this scenario should be considered as might the attacker don't have complete control of the communication channel and can only add slight perturbs.

In the fourth point, the attack can also happen if the system (example: server) is compromised. This scenario, also similar to the third point. It more of an intrusion in the system. Low perturbs attack can happen, but we can see that if the attacker already compromised the server, attacker doesn't need to compromise the input. It is easier for attacker to change the output label, as attacker already compromised the system. We can argue that the attacker's ability is limited, so that attacker will prefer AAs.

The fifth point is very different compare to other points, and this attack can happen before the ML model has been set up on the server. Some Trojans or backdoor can be added in the ML before deployment, which is only activated for specific inputs. 

Additional point which is not presented in the figure was adaptive attack. An attacker can run different test input to determine the decision boundary of ML model. Also, as AAs are transferable attack generated in another system will work in here. 

From our above discussion, it is evident that AAs can happen outside the physical world that can make an ML model provide wrong output.
\section{Who will Enforce the Defense?}
This section will discuss which domain of expertise is required to defend against the adversarial threat. Based on our discussion in section \ref{ch} we can see adversarial defenses each have a different implementation. For example, adversarial training or distillation technique requires training the ML model, resulting in data scientist purview. A prepossessing approach such as feature squeezing types of defense may employ by system engineering in run-time. But the limitations of these defense techniques create a challenge for the concerned parties. Pre-processing methods can protect most of the standard AA, and it can employ when input sensors are converting to digital format. 

Second point attack, as described in figure \ref{fig:radar} may concern with knowledge concerning hardware security or IoT device security, or firmware security. This part interacts with humans, so human-computer interaction-based defenses (trust models?) may fall in this area. Pre-processing-based adversarial defense technique can be employed here before sending over the communication channel.

Third point attacks, as described in figure \ref{fig:radar} are in the communication channel, so an attack on this channel should be concerned by network security experts. As if this channel is compromised, an attacker can create different attacks rather than AAs. Some of these attacks may be a more severe threat than AAs. Computer network layer security, data layer security also concerned here. It is in question that is it possible to use any of the adversarial defense technique can be employed in this stage or not. It should be noted that traditional adversarial defenses, as far as our literature review, are not applicable in this area, and here are new research opportunities for the interested cybersecurity experts as they can explore from AA perspectives with traditional attack types.

The fourth point directly falls in operating system security or cloud security based ML implementation. It could be argued that prepossess-based defense technique can also be employed here, but if there is an intrusion occurs, that we have more vital concerns than adversarial attacks as the system is already compromised. The attacker is effectively controlling the system. Many proposed adversarial defense techniques can be implemented here. Still, all of these have a limitation: an attacker can modify data in the OS label. Attacker can probably corrupt the employed defense system there too. 

The fifth point of attack can be defended by checking the output with other AI systems or conducting thorough testing. Effectively as the ML model is considered as a black box system, an engineer or cybersecurity analyst has a minimal role to play. Data scientists can develop some algorithms to test the ML model after deployment to discover unexpected results if exist. We consider this to be the most concerning attack points as it is hard to address and may need extensive  testing on a regular basis.

The adaptive attack is another grave concern, which is also hard to defend as adversarial inputs are transferable. However, it can be handled with dynamic ML model configuration (with regular update), or using active/reinforcement learning approach. It can also help if employ a defense system to identify attack query pattern, which can relate to DDoS attacks. However, one downside of adaptive attack is it is a computationally expensive and time-consuming for a system that is regularly updated.
\begin{table}[h]
\centering
\tiny
\begin{tabular}{|l|l|l|}
\hline
\multicolumn{1}{|c|}{\textbf{Attack point}} & \multicolumn{1}{c|}{\textbf{Defense}} & \multicolumn{1}{c|}{\textbf{Role}} \\ \hline
Input Modification                          & Preprocess technique                  & System/Software devs     \\ \hline
Sensor,I/O device                           & Testing and Reliability               & Devops /AIops/MLops                            \\ \hline
Communication Channel                       & Network defense, Cryptography         & Network and IT security      \\ \hline
Internal OS system                          & Intrusion detection, Access control   & Network and IT security      \\ \hline
Model Structure                             & Formal methods                               & Data/ML scientist                     \\ \hline
Adaptive attack                             & Dynamic and active learning           & Data/ML scientist/Network          \\ \hline
\end{tabular}
\caption{Different attack points and relevant responsible professional}
\label{tab:my-table}
\end{table}
From the above discussion, the table \ref{tab:my-table} is developed, where we illustrated what part of defense falls in which security domain. This table is an attempt to divide the roles among AI/ML engineers and administrators, we also  assume that there need to be more intercommunication among different roles to implement a robust defense system. 

It is evident that in the literature survey provided in section \ref{ch}, most of the defense strategies except prepossessing techniques are not suitable in deployed environments. Developing an adversarial defense technique that can work in deployed environment and cover all the attack points remain an open challenge to researchers.

\section{Key Takeaway}
Adversarial attack (AA) formulations are effective for studying and understanding how deep learning methods work, and it has enormous potential to devise a way to provide explainability in AI. But as a cybersecurity threat, it's not yet a considerable threat that would require significant cyber resources other than the standard cybersecurity practices. But it cannot be ignored that as more research progresses, it will become a substantial threat that would cause the AI-based system vulnerable. Considering the current technical and economic dependency of AI-based systems, it would be disastrous if we don't address this issue and keep ML systems exposed to AAs. We need a coordinated defense planning with MLOps, System Admins, Software engineers, cyber experts.

\section{Closure}
We can conclude that adversarial defenses are closely intertwined with other application-specific security domains. Most adversarial defense techniques are devised without considering these aspects, limiting their usefulness in different implementation. An adversarial defense focused on training the ML models or changing the architectures, and is not applicable for deployed systems. Most of the AA's fail to work in the physical environment also raises questions about evaluating a defense system as the current adversarial defense bench-marking system prioritizes performance against maximum attack types. Defense against adaptive attack remains the hardest challenge, but it is also evident formulate an adaptive attack is computationally expensive if the system has a dynamic ensemble of active learning approaches. In short, our above discussion  can summarize in the below points
\begin{itemize}
\small
    \item Adversarial defenses are interconnected with other cybersecurity and network infrastructure domains.
    \item There was a lack of research for adversarial defenses in deployed environments, and need higher priorities before releasing AI-based products to market.
    \item Adaptive attacks and TrojAI attacks are the most challenging which will require end-to-end protection to defend.
    \item Defense against AAs needs a collaborative approach between network, data, firmware, hardware, OS, cloud security.
\end{itemize}
In this paper, we generalized complex concepts of adversarial defenses in the most straightforward terms to ease discussion. We believe this discussion will motivate the research community to comprehend the deployed AI/ML model security from broader perspectives. 


\bibliography{example_paper}
\bibliographystyle{icml2021}

\end{document}